\documentclass[prb,preprint,showpacs,preprintnumbers,amsmath,amssymb]{revtex4}
\usepackage{bm}
\usepackage{epsfig}

\begin{document}
\title{Two-time quantum transport and quantum diffusion}
\author{P. Kleinert}
\email[]{kl@pdi-berlin.de}
\affiliation{Paul-Drude-Intitut f\"ur Festk\"orperelektronik,
Hausvogteiplatz 5-7, 10117 Berlin, Germany}
\date{\today}
\begin{abstract}
Based on the nonequilibrium Green's function technique, a unified theory is developed that covers quantum transport and quantum diffusion in bulk semiconductors on the same footing. This approach, which is applicable to transport via extended and localized states, extends previous semi-phenomenological studies and puts them on a firm microscopic basis. The approach is sufficiently general and applies not only to well-studied quantum transport problems, but also to models, in which the Hamiltonian does not commute with the dipole operator. It is shown that, even for the unified treatment of quantum transport and quantum diffusion in homogeneous systems, all quasi-momenta of the carrier distribution function are present and fulfill their specific function. Particular emphasis is put on the double-time nature of quantum kinetics. To demonstrate the existence of robust macroscopic transport effects that have a true double-time character, a phononless steady-state current is identified that appears only beyond the generalized Kadanoff-Baym ansatz.
\end{abstract}


\maketitle

\section{Introduction}
Quantum transport in semiconductors has attracted a great deal of interest over the last few decades. The transport theory of carriers in bulk semiconductors as developed on the basis of semiclassical Boltzmann or balance equations as well as the nonequilibrium Green's function (GF) technique is well documented in many review articles and textbooks (cf., for instance, Refs. [\onlinecite{Reggiani,Haug,Ferry}]) and has reached a high level of sophistication. Compared with this achievement, there are only a few studies on quantum diffusion in semiconductors. However, this fact does not imply that the knowledge about quantum diffusion has not contributed to the overall picture of quantum kinetics. It is rather the unified description of ballistic transport, diffusion, and hopping that provides valuable insight into the general structure of the theory. This paper aims at the construction of a general transport theory by focusing on common features that emerge from quantum transport and quantum diffusion as well as from transport via extended and localized states. A comparative analysis of quantum transport and quantum diffusion, carried out on the basis of a semi-phenomenological approach, \cite{Bryksin:1415,Kleinert:315} clearly stressed distinct features of the general theory that already appeared in early studies of the current density. \cite{Bryksin:3344} The problem is best explained by treating the momentum representation of the one-particle transport approach. It is usually assumed that due to translational invariance the carrier transport in homogeneous systems is properly accounted for by only one ${\bm k}$ vector. Consequently, the carrier transport should be governed by the distribution function $f^{<}({\bm k}|t)$, which is the solution of a quantum-kinetic equation. However, completely different results were obtained from a unified approach to quantum transport and quantum diffusion on the basis of the conditional transition probability that satisfies a quantum-mechanical Bethe-Salpeter equation. \cite{Bryksin:1415,Kleinert:315} The main quantity in this approach is the distribution function $f^{<}({\bm k},{\bm\kappa}|t)$ that depends on both wave vectors, although the underlying model is completely homogeneous and not strongly affected by special initial conditions. This finding is all the more surprising since the vector ${\bm\kappa}$ refers to a deviation from homogeneity. In fact, it is not the full ${\bm\kappa}$ dependence  that is needed in the calculation of transport coefficients of homogeneous bulk semiconductors. What enters the approach is the quantity $\nabla_{\bm\kappa}f^{<}({\bm k},{\bm \kappa}|t)|_{{\bm\kappa}={\bm 0}}$, which is interpreted to be a virtual disturbance of the homogeneous system that allows to probe the drift-diffusion response. The surprising result that both wave vectors ${\bm k}$ and ${\bm\kappa}$ appear also in the general transport theory of homogeneous systems is confirmed by the unified picture of transport via extended and localized states. In principle, hopping and band transport can be described on the same footing so that each approach is derivable from the other one in a straightforward manner. This equivalence is of particular importance, when transport in biased superlattices is treated. By tuning the electric field applied perpendicular to the layers of the superlattice, the character of transport can be driven from hopping to the Ohmic regime and vice versa. From an application point of view, however, it is natural to expect that both approaches are not equally well adapted for numerical purposes.

To provide further arguments for the statement that also the vector ${\bm\kappa}$ enters the general transport theory, let us focus on the current density ${\bm j}(t)$, which is given by the time derivative of the dipole operator ${\bm j}(t)=(1/V)d{\bm D}(t)/dt$ (with $V$ being the volume of the system). In the momentum representation, an equivalent form applicable to the steady state is obtained by
\begin{equation}
{\bm j}=\sum\limits_{\bm k}{\bm v}_{eff}({\bm k})f^{<}({\bm k}),
\label{Intro}
\end{equation}
where ${\bm v}_{eff}({\bm k})$ denotes an effective velocity. In the majority of transport studies, ${\bm v}_{eff}({\bm k})$ is simply given by the drift velocity ${\bm v}({\bm k})=\nabla_{\bm k}\varepsilon({\bm k})/\hbar$ that refers to extended states with the kinetic energy $\varepsilon({\bm k})$. This result applies, whenever the interaction Hamiltonian $H_{int}$ commutes with the dipole operator, which, however, does not always happen. For instance, in the theory of small polarons, there is an extra current contribution ${\bm j}\sim [{\bm D},H_{int}]_{-}$, which even dominates the carrier transport via the effective drift velocity ${\bm v}_{eff}({\bm k})$. The current density of this kind is expressed by the ${\bm\kappa}$ derivative of the collision integral calculated at ${\bm\kappa}={\bm 0}$. Another example refers to the proper definition of the spin current in semiconductors with spin-orbit interaction that recently became a subject of a lively controversial discussion. \cite{wir_Spinstrom} In addition, the exact switching from band to hopping transport and vice versa requires the consideration of both momenta ${\bm k}$ and ${\bm\kappa}$ in the scattering terms and GFs. \cite{Bryksin:1415}

The former semi-phenomenological studies provided a unified approach to quantum transport and quantum diffusion that covers transport via localized and extended states. \cite{Bryksin:1415,Kleinert:315} It remains the task to put this suggestive transport picture on a firm microscopic foundation. The most attractive starting point for this purpose provides the nonequilibrium GF technique that was initiated by Schwinger \cite{Schwinger} and Keldysh \cite{Keldysh} and that has been employed by several authors \cite{Korenman,Danielewicz,Chou_PhysRep,Rammer,Botermans,Zubarev} to treat quantum transport. With respect to the time dependence, such a fundamental approach is able to cope with the two-time nature of quantum evolution. The double-time character manifests itself in coupled evolution equations for the spectral function and the statistical propagator. Moreover, memory effects appear due to the integration over the full time history. Unfortunately, most applications of the nonequilibrium GF method disregarded the entangled time dependence by relying on a sufficient homogeneity in time, which is exploited by first-order gradient expansions, when 'center of mass' coordinates in space and time slowly vary. This approximation has a serious disadvantage, namely it rules out the full quantum-mechanical character of transport phenomena. In contrast to these approaches, we account for the two-time dependence of GFs in an exact manner and generalize our former unified approach to quantum transport and quantum diffusion. \cite{Bryksin:1415} The basic quantity will be the two-time distribution function $f^{<}({\bm k},{\bm\kappa}|T,t)$ that satisfies a quantum-kinetic equation. An example, given in the last Section, demonstrates the existence of a macroscopic phononless quantum transport mechanism that emerges only, when the double-time dependence of GFs is properly taken into account.

\section{Basic approach}
\subsection{Symmetry of the Green functions}
We are going to focus on the basic physics of quantum transport that is revealed by controlled approximations in a one-particle picture that disregards the Coulomb interaction between carriers. Consequently, the electron-phonon interaction takes over the indispensable role of an inelastic scattering mechanism, which is needed when treating the nonlinear high-field transport. In spite of this restriction, we will, nevertheless, consider a rather general model that is compatible with many studies of quantum transport in semiconductors.

The main issue in deriving basic quantum-kinetic equations is the full account of symmetries. Although the translational invariance with respect to the spatial ${\bm r}_{i}$ and temporal $t_{i}$ coordinates is broken, when external electric ${\bm E}(t)$ and magnetic ${\bm B}$ fields are applied to the sample, there remains a symmetry of GFs that is very important for the description of field-dependent nonequilibrium quantum transport. \cite{Kleinert:4123} This symmetry expresses the fact that a translation of spatial coordinates can be compensated by the vector potentials ${\bm A}({\bm r})$ and ${\bm A}(t)$ of the external magnetic and time-dependent electric field, respectively. For the expectation values of the one-particle propagators $G^{\gtrless }$, we have
\begin{equation}
G^{\gtrless }({\bm r}_1,t_1| {\bm r}_2,t_2)=e^{i{\bm A}(t_2,t_1){\bm r}
+i{\bm A}({\bm r})({\bm r}_2-{\bm r}_1)}
G^{\gtrless }({\bm r}_1+{\bm r},t_1| {\bm r}_2+{\bm r},t_2),
\label{Seq1}
\end{equation}
with the abbreviations ${\bm A}(t_2,t_1)={\bm A}(t_2)-{\bm A}(t_1)$ and $d{\bm A}(t)/dt=e{\bm E}(t)/\hbar$, where the vector potential of the magnetic field ${\bm A}({\bm r})$ is given in the symmetric gauge. The symmetry expressed by Eq.~(\ref{Seq1}) favors the utilization of the so-called Wigner transformed GFs denoted by $\widetilde{G}^{\gtrless }$ that are invariant under spatial translations
\begin{equation}
G^{\gtrless }({\bm r}_1,t_1| {\bm r}_2,t_2)
=\widetilde{G}^{\gtrless }({\bm r}_1,t_1| {\bm r}_2,t_2)
e^{i{\bm A}({\bm r}_2){\bm r}_1-i{\bm A}(t_2,t_1)({\bm r}_1+{\bm r}_2)/2},
\label{Seq2}
\end{equation}
with
\begin{equation}
\widetilde{G}^{\gtrless }({\bm r}_1+{\bm r},t_1| {\bm r}_2+{\bm r},t_2)
=\widetilde{G}^{\gtrless }({\bm r}_1,t_1| {\bm r}_2,t_2).
\label{Seq3}
\end{equation}
For a constant electric field, $\widetilde{G}^{\gtrless}$ agree with gauge-invariant GFs (cf., for instance, Ref. [\onlinecite{Levanda:7889,Kremp:4725}]). Equation (\ref{Seq2}) leads to simplifications that are most effectively exploited in Fourier space. Changing the coordinates according to
\begin{equation}
{\bm R}=\frac{{\bm r}_1+{\bm r}_2}{2},\quad
{\bm r}={\bm r}_2-{\bm r}_1,\quad
T=\frac{t_1+t_2}{2},\quad
t=t_2-t_1,
\label{Seq4}
\end{equation}
we perform a Fourier transformation with respect to the spatial coordinates ${\bm r}$ and ${\bm R}$ to obtain
\begin{equation}
G^{\gtrless }({\bm k},{\bm\kappa}|T,t)=
\int d^{3}{\bm r} e^{-i({\bm\kappa}-{\bm A}(T,t)){\bm r}}
\widetilde{G}^{\gtrless }({\bm k}+\frac{{\bm\kappa}}{2}
-\frac{1}{2}{\bm A}(T,t)-{\bm A}({\bm r})|T,t),
\label{Seq5}
\end{equation}
with
\begin{equation}
{\bm A}(T,t)={\bm A}\left(T-\frac{t}{2}\right)-{\bm A}\left(T+\frac{t}{2}\right).
\label{Seq6}
\end{equation}
The reduction of degrees of freedom and the separation of the momentum ${\bm\kappa}$ become more transparent in the absence of a magnetic field, when Eq.~(\ref{Seq5}) is converted to the form
\begin{equation}
G^{\gtrless }({\bm k},{\bm\kappa}|T,t)=\delta({\bm\kappa}-{\bm A}(T,t))
\widetilde{G}^{\gtrless }({\bm k}|T,t).
\label{Seq7}
\end{equation}
This equation enables the formulation of the transport theory on the basis of GFs $\widetilde{G}^{\gtrless}$ that respect the symmetry in the presence of electromagnetic fields. In our former quantum approach to carrier transport, \cite{Bryksin:233,Bryksin:1235} we profited from this transformation and from the general symmetry relation
\begin{equation}
G^{\gtrless }({\bm k},{\bm\kappa}|T,t)^{*}=
-G^{\gtrless }({\bm k},-{\bm\kappa}|T,-t).
\label{Seq8}
\end{equation}
Note that the wave vector ${\bm\kappa}$ refers to a possible deviation from homogeneity and is absent in the conventional treatment of transport in homogeneous systems.

\subsection{Dyson equation}
A nonequilibrium system is completely characterized by only two independent two-point functions. From a physical point of view, the decomposition of the full matrix of GFs into statistical and spectral components is most attractive due to its clear physical interpretation and due to the unambiguous separation of their dynamical role. Loosely speaking, this choice of GFs makes clear which states are available and how often they are occupied. Having this natural decomposition in mind, we derive kinetic equations for the GFs $G^{>}$ and $G^{<}$, which effectively yield the density of states and the double-time density matrix, respectively. In the momentum representation, the coupled Dyson equations have the form \cite{Bryksin:2731}
\begin{eqnarray}
&&\left[ i\hbar \frac{\partial }{\partial t}-\varepsilon ({\bm k})+ie{\bm E}%
(t)\nabla _{{\bm k}}\right] G^{\gtrless }({\bm k}t\mid {\bm k}^{\prime
}t^{\prime })  \nonumber \\
&=&\pm \hbar \int d{\bm k}_{1}\biggl\{ \int\limits_{t^{\prime
}}^{t}dt_{1}\Sigma ^{\gtrless }({\bm k}t\mid {\bm k}_{1}t_{1})G^{\gtrless }(%
{\bm k}_{1}t_{1}\mid {\bm k}^{\prime }t^{\prime })  \nonumber \\
&+&\int\limits_{-\infty }^{t^{\prime }}dt_{1}\Sigma ^{\gtrless }({\bm k}%
t\mid {\bm k}_{1}t_{1})G^{\lessgtr }({\bm k}_{1}t_{1}\mid {\bm k}%
^{\prime}t^{\prime })  \nonumber \\
&-&\int\limits_{-\infty }^{t}dt_{1}\Sigma ^{\lessgtr } ({\bm k}t\mid {\bm k}%
_{1}t_{1})G^{\gtrless }({\bm k}_{1}t_{1}\mid {\bm k}^{\prime}t^{\prime })%
\biggl\}.
\label{Dyson1}
\end{eqnarray}
All scattering contributions are included in the self-energies $\Sigma ^{\lessgtr }$, while the time-dependent electric field is treated in the vector potential gauge. The derivation of kinetic equations for the GFs proceeds by well established steps: (i) new wave vectors are introduced by the replacement ${\bm k}\rightarrow {\bm k}+{\bm\kappa}/2$, ${\bm k}^{\prime}\rightarrow {\bm k}-{\bm\kappa}/2$ and (ii) Dyson equations written down for $G^{\gtrless }({\bm k},{\bm\kappa}|T,t)$ and $G^{\gtrless }({\bm k},{\bm\kappa}|T,-t)^{*}$ are subtracted from each other. This procedure leads to exact quantum-kinetic equations. For simplicity, let us shorten the cumbersome calculation by focusing on a non-degenerate electron gas, for which the Boltzmann statistics applies under idealized conditions. As a consequence of this simplification, there is a strong imbalance between $G^{<}$ and $G^{>}$ in the sense that $G^{<}$ is much 'smaller' than $G^{>}$ as the statistical propagator $G^{<}$ is proportional to the carrier density. Adopting this approximation, we arrive at the following non-Markovian integral equation for the correlation function $G^{<}$
\begin{eqnarray}
&&\left[ i\hbar \frac{\partial }{\partial T}+\varepsilon({\bm k}-\frac{{\bm\kappa}}{2})
-\varepsilon({\bm k}+\frac{{\bm\kappa}}{2})-i\hbar\frac{\partial {\bm A}(T,t)}{\partial t}\nabla_{\bm k}
+i\hbar\frac{\partial {\bm A}(T,t)}{\partial T}\nabla_{\bm\kappa}\right]
G^{<}({\bm k},{\bm\kappa}|T,t)\nonumber\\
&&=\hbar\sum\limits_{\bm\kappa_1}\int\limits_0^{\infty}dt_1\biggl\{
-\Sigma^{<}({\bm k}+\frac{{\bm\kappa}_1}{2},{\bm\kappa}-{\bm\kappa}_1|
T-\frac{t_1}{2},t-t_1)
G^{>}({\bm k}-\frac{{\bm\kappa}-{\bm\kappa}_1}{2},{\bm\kappa}_1|
T+\frac{t-t_1}{2},t_1)\nonumber\\
&&-\Sigma^{<}({\bm k}-\frac{{\bm\kappa}_1}{2},{\bm\kappa}-{\bm\kappa}_1|
T-\frac{t_1}{2},t+t_1)
G^{>}({\bm k}+\frac{{\bm\kappa}-{\bm\kappa}_1}{2},{\bm\kappa}_1|
T-\frac{t+t_1}{2},-t_1)\nonumber\\
&&+G^{<}({\bm k}-\frac{{\bm\kappa}_1}{2},{\bm\kappa}-{\bm\kappa}_1|
T-\frac{t_1}{2},t+t_1)
\Sigma^{>}({\bm k}+\frac{{\bm\kappa}-{\bm\kappa}_1}{2},{\bm\kappa}_1|
T-\frac{t+t_1}{2},-t_1)\nonumber\\
&&+G^{<}({\bm k}+\frac{{\bm\kappa}_1}{2},{\bm\kappa}-{\bm\kappa}_1|
T-\frac{t_1}{2},t-t_1)
\Sigma^{>}({\bm k}-\frac{{\bm\kappa}-{\bm\kappa}_1}{2},{\bm\kappa}_1|
T+\frac{t-t_1}{2},t_1)\biggl\}.
\label{Dyson2}
\end{eqnarray}
To proceed further, one has to specify expressions for the self energies, which are dictated by the underlying model (e.g., small polarons), the treated scattering diagrams (e.g., $T$ matrix approximation), and the possible inclusion of the initial conditions. \cite{Semkat:7458} As an example, let us select the self-consistent Born approximation with coupling functions $U_{\lambda}^{\gtrless}$ that are local in time
\begin{equation}
\Sigma^{\gtrless }({\bm k},{\bm\kappa}|T,t)=\sum\limits_{{\bm k}_1,\lambda}
U^{\gtrless }_{\lambda}({\bm k},{\bm k}_1,{\bm\kappa}|T,t)
G^{\gtrless }({\bm k}_1,{\bm\kappa}|T,t).
\label{Born}
\end{equation}
For the widespread Fr\"ohlich-type electron-phonon interaction, this equation reduces to
\begin{equation}
U^{\gtrless }_{\lambda}({\bm k},{\bm k}^{\prime},{\bm\kappa}|T,t)=
D^{\gtrless }_{\lambda}({\bm k}^{\prime}-{\bm k}|t).
\end{equation}
The kinetic Eq.~(\ref{Dyson2}) together with Eq.~(\ref{Born}) are not in the form that is commonly used for the calculation of the current density and the diffusion coefficient. Further transformation steps have to be carried out without any approximation. An obvious procedure would be the full exploitation of the symmetry based on the transformation in Eq.~(\ref{Seq5}), which would considerably simplify the approach to nonequilibrium quantum transport. A consequent exploitation of the Wigner-transformed GFs $\widetilde{G}^{\gtrless}$, as applied in our former approach, \cite{Bryksin:2731,Bryksin:233,Bryksin:1235} leads, however, to a complete disappearance of the wave vector ${\bm\kappa}$ that plays a fundamental role in the unified quantum transport description of both carrier drift and diffusion. Therefore, we suggest another procedure, which is more general and which discriminates between the states of the system ($G^{>}$) and their dynamical evolution ($G^{<}$). The clear distinction between the role of spectral and dynamical GFs, which is a generic feature of nonequilibrium field theory, is accounted for in the new approach by maintaining the GF $G^{<}$ in Eq.~(\ref{Dyson2}) in its original form and by replacing only the GF $G^{>}$ by the symmetry-adapted partner $\widetilde{G}^{>}$ defined in Eq.~(\ref{Seq7}). Loosely speaking, the remaining ${\bm\kappa}$ dependence in $G^{<}$ is needed to simulate a virtual probe of the system that reveals its dynamical response.

The reformulation of the kinetic Eq.~(\ref{Dyson2}) is facilitated by introducing new functions $\overline{G}^{<}$ and $\overline{U}_{\lambda}^{\gtrless}$
\begin{equation}
G^{<}({\bm k},{\bm\kappa}|T,t)=
\overline{G}^{<}({\bm k},{\bm\kappa}-{\bm A}(T,t)|T,t),
\label{Gquer}
\end{equation}
\begin{equation}
U^{\gtrless }_{\lambda}({\bm k},{\bm k}_1,{\bm\kappa}|T,t)=
\overline{U}^{\gtrless }_{\lambda}({\bm k},{\bm k}_1,{\bm\kappa}-{\bm A}(T,t)|T,t),
\label{Uquer}
\end{equation}
that account for the internal order of the kinetic equations with respect to ${\bm\kappa}$. By a further transformation, the kinetic energy of carriers is separated out by replacing the basic GFs $\widetilde{G}^{>}$ and $\overline{G}^{<}$ through new ones $R^{>}$ and $R^{<}$, which are defined by
\begin{equation}
\widetilde{G}^{>}({\bm k}|T,t)=-iR^{>}({\bm k}|T,t)\exp\biggl\{
\frac{i}{\hbar}\int_{-t/2}^{t/2}d\tau\varepsilon
\left({\bm k}+{\bm A}(T,t)-\frac{1}{2}\left[{\bm A}(T+\frac{t}{2})+{\bm A}(T-\frac{t}{2})\right]\right)
\biggl\},
\label{Rg}
\end{equation}
\begin{eqnarray}
&&\overline{G}^{<}({\bm k},{\bm\kappa}|T,t)=iR^{<}({\bm k},{\bm\kappa}|T,t)
\label{Rk}\\
&&\exp\biggl\{\frac{i}{2\hbar}\int\limits_{-t/2}^{t/2}d\tau\left[
\varepsilon\left({\bm k}-\frac{\bm\kappa}{2}+{\bm A}(T+\tau)-\frac{1}{2}
\left[{\bm A}(T+\frac{t}{2})+{\bm A}(T-\frac{t}{2})\right]\right)
\right.\nonumber\\
&&\left. +
\varepsilon\left({\bm k}+\frac{\bm\kappa}{2}+{\bm A}(T+\tau)-\frac{1}{2}
\left[{\bm A}(T+\frac{t}{2})+{\bm A}(T-\frac{t}{2})\right]\right)
\right]
\biggl\}.\nonumber
\end{eqnarray}
The reformulation of the kinetic equation, which is based on Eqs.~(\ref{Rg}) and (\ref{Rk}), isolates rapidly varying phase factors and leads to the conventional collision integral that appears in transport theory. The final exact reconstruction of the original kinetic Eq.~(\ref{Dyson2}) is carried out by introducing the physical GFs $f^{\gtrless}$ and respective coupling terms ${\cal U}_{\lambda}^{\gtrless}$ via
\begin{equation}
R^{\gtrless}({\bm k},{\bm\kappa}|T,t)=f^{\gtrless}({\bm k}\pm\frac{1}{2}{\bm A}(T,t),{\bm\kappa}|T,t),
\quad {\rm for}\quad t \gtrless 0,
\label{fkg}
\end{equation}
\begin{equation}
\overline{U}^{\gtrless}_{\lambda}({\bm k},{\bm k}^{\prime},{\bm\kappa}|T,t)=
{\cal U}^{\gtrless}_{\lambda}({\bm k}\pm\frac{1}{2}{\bm A}(T,t),
{\bm k}^{\prime}\pm\frac{1}{2}{\bm A}(T,t),{\bm\kappa}|T,t),
\quad {\rm for}\quad t \gtrless 0.
\label{schonU}
\end{equation}
The execution of all transformation steps is straightforward and leads to a final kinetic equation, in which a time-dependent renormalization of the bare kinetic energy appears on the left-hand side. As these single-particle corrections disappear in the limit $t\rightarrow 0$, which is relevant for the calculation of all transport coefficients, we neglect these contributions. This approximation does not significantly affect the general character of our approach. Inserting Eqs.~(\ref{Gquer}) to (\ref{schonU}) into the kinetic Eq.~(\ref{Dyson2}), we obtain our main general result, namely a quantum-kinetic equation for the carrier distribution function $f^{<}$ that characterizes the statistical properties of the nonequilibrium system
\begin{eqnarray}
&&\left\{\frac{\partial }{\partial T}+\frac{i}{\hbar}\left(\varepsilon({\bm k}+\frac{{\bm\kappa}}{2})
-\varepsilon({\bm k}-\frac{{\bm\kappa}}{2})\right)-\frac{\partial {\bm A}(T,t)}{\partial t}\nabla_{\bm k}
+\frac{\partial {\bm A}(T,t)}{\partial T}\nabla_{\bm\kappa}\right\}
f^{<}({\bm k},{\bm\kappa}|T,t)\label{Dyson3}\\
&&=\sum\limits_{{\bm q},\lambda}\sum\limits_{{\bm k}_1}\biggl\{
\int\limits_{0}^{\infty}dt_1\left[\delta_{{\bm k}_1,{\bm k}+{\bm q}+\Delta{\bm A}_{+}}
{\cal U}_{\lambda}^{<}({\bm k}_1-{\bm q},{\bm k}_1,{\bm\kappa}|T-\frac{t_1}{2},t-t_1)
\right.\nonumber\\
&&\left. -\delta_{{\bm k}_1,{\bm k}+\Delta{\bm A}_{+}}
{\cal U}_{\lambda}^{>}({\bm k}_1-\frac{{\bm\kappa}}{2},{\bm k}_1-{\bm q}-\frac{\bm{\kappa}}{2},{\bm 0}|T+\frac{t-t_1}{2},t_1) \right]\nonumber\\
&&\times f^{<}({\bm k}_1,{\bm\kappa}|T-\frac{t_1}{2},t-t_1)
f^{>}({\bm k}_1-{\bm q}-\frac{\bm\kappa}{2}|T+\frac{t-t_1}{2},t_1)
P_{t_1}({\bm k}_1,{\bm q},{\bm\kappa}|T,t)\nonumber\\
&&-\int\limits_{-\infty}^{0}dt_1\left[\delta_{{\bm k}_1,{\bm k}+\Delta{\bm A}_{-}}
{\cal U}_{\lambda}^{>}({\bm k}_1+\frac{{\bm\kappa}}{2},{\bm k}_1-{\bm q}+\frac{\bm\kappa}{2},{\bm 0}|T-\frac{t-t_1}{2},t_1)
\right.\nonumber\\
&&\left. -\delta_{{\bm k}_1,{\bm k}+{\bm q}+\Delta{\bm A}_{-}}
{\cal U}_{\lambda}^{<}({\bm k}_1-{\bm q},{\bm k}_1,{\bm\kappa}|T+\frac{t_1}{2},t-t_1) \right]\nonumber\\
&&\times f^{<}({\bm k}_1,{\bm\kappa}|T+\frac{t_1}{2},t-t_1)
f^{>}({\bm k}_1-{\bm q}+\frac{\bm\kappa}{2}|T-\frac{t-t_1}{2},t_1)
Q_{t_1}({\bm k}_1,{\bm q},{\bm\kappa}|T,t)
\biggl\}.\nonumber
\end{eqnarray}
The quantities that appear in this kinetic equation are given by
\begin{equation}
\Delta{\bm A}_{+}={\bm A}(T+\frac{t}{2}-t_1)-{\bm A}(T-\frac{t}{2}),\quad
\Delta{\bm A}_{-}={\bm A}(T-\frac{t}{2}+t_1)-{\bm A}(T-\frac{t}{2}),
\end{equation}
\begin{equation}
P_{t_1}({\bm k}_1,{\bm q},{\bm\kappa}|T,t)=
F_{t}({\bm k},{\bm\kappa}|T-\frac{t}{2})
F_{t_1}({\bm k}_1,{\bm q},{\bm\kappa}|T+\frac{t}{2}-t_1)
F_{t_1-t}({\bm k}_1,{\bm \kappa}|T+\frac{t}{2}-t_1),
\label{eqP}
\end{equation}
\begin{equation}
Q_{t_1}({\bm k}_1,{\bm q},{\bm\kappa}|T,t)=
F_{t}({\bm k},{\bm\kappa}|T-\frac{t}{2})
F_{-t_1}^{*}({\bm k}_1,{\bm q},-{\bm\kappa}|T-\frac{t}{2}+t_1)
F_{-(t_1-t)}^{*}({\bm k}_1,{\bm \kappa}|T-\frac{t}{2}+t_1),
\label{eqQ}
\end{equation}
\begin{equation}
F_{t_1}({\bm k},{\bm q},{\bm\kappa}|t_2)=
\exp\left\{\frac{i}{\hbar}\int\limits_{0}^{t_1}d\tau
\varepsilon\left({\bm k}-{\bm q}-\frac{\bm\kappa}{2}+{\bm A}(t_2+\tau)
-{\bm A}(t_2) \right) \right\},
\label{FF1}
\end{equation}
\begin{eqnarray}
F_{t_1}({\bm k},{\bm\kappa}|t_2)&&=
\exp\biggl\{-\frac{i}{2\hbar}\int\limits_{0}^{t_1}d\tau\left[
\varepsilon\left({\bm k}-\frac{\bm\kappa}{2}+{\bm A}(t_2+\tau)
-{\bm A}(t_2) \right)\right.\label{FF2}\\
&&\left. +
\varepsilon\left({\bm k}+\frac{\bm\kappa}{2}+{\bm A}(t_2+\tau)
-{\bm A}(t_2) \right)\right] \biggl\}.\nonumber
\end{eqnarray}
The general result in Eq.~(\ref{Dyson3}) has a number of intriguing features. First of all, we mention its two-time character. All quantities that enter the kinetic Eq.~(\ref{Dyson3}) depend on two time variables that are responsible for the evolution on different time scales. This generic double-time nature of the nonequilibrium dynamics manifests itself in memory effects, \cite{Hartmann:165} which are revealed by the non-Markovian time dependence in the collision integral. In general, the time convolution occurs in both time domains, but it is assumed that the most prominent effect primarily happens on the microscopic time scale. For the calculation of transport coefficients, only the special one-time distribution function $f^{<}({\bm k},{\bm\kappa}|T,t=0)$ is needed. However, its determination from Eq.~(\ref{Dyson3}) is still confronted with the double-time nature of the problem that resides in the right-hand-side (RHS) of this equation. A solution of this reconstruction problem is offered by the generalized Kadanoff-Baym ansatz. \cite{Lipavski:6933,Velicky:041201} But in this approximation, the two-time dependence is notoriously discarded. The physics that emerges beyond this approximation is captured by maintaining the general double-time quantum-kinetic Eq.~(\ref{Dyson3}).

Let us add a general remark concerning the interpretation of the double-time kinetics. As shown in the next Subsection, the distribution function $f^{<}({\bm k},{\bm\kappa}|T,t=0)$ fully determines the temporal order of macroscopic transport phenomena. However, this quantity is only given by a strict reference to its two-time extension. Moreover, the self-energy is generally constructed from many-particle GFs that depend on several time variables. Therefore, on a fundamental level, the change of events cannot be brought into a sequence that can be mapped on a one-parameter flow of time extending from the past to the future. In quantum statistics, the changeability of things seems to be a more general concept than the temporal evolution based on an absolute time in the sense of classical physics.

Most essential for the construction of a unified kinetic theory that covers both the carrier drift and diffusion is the presence of the second wave vector ${\bm\kappa}$ in Eq.~(\ref{Dyson3}), which, generally speaking, refers to a spatial inhomogeneity. Similar to the role played by the time variable $t$, the full ${\bm\kappa}$ dependence does not enter the expressions for the transport quantities, but only ${\bm\kappa}$ gradients at ${\bm\kappa}={\bm 0}$ (for instance, the basic quantity for quantum diffusion \cite{Bryksin:1415} is given by the vector $\nabla_{\bm\kappa}f^{<}({\bm k},{\bm\kappa}|T,0)|_{{\bm\kappa}={\bm 0}}$). Besides the doubling of spatial and time variables, we mention an additional peculiarity of the quantum-kinetic Eq.~(\ref{Dyson3}), namely the shift of momentum variables in Eqs.~(\ref{FF1}) and (\ref{FF2}) by the vector potential of the electric field. This explicit field dependence of scattering gives rise to intracollisional field effects and nonlinear transport.

At the end of this Subsection, we will show that the general two-time quantum-kinetic Eq.~(\ref{Dyson3}) reproduces a number of established results that appear, when further assumptions are adopted. The most familiar form of kinetic equations is obtained, when the macroscopic time scale is inert to microscopic fluctuations ($T+\Delta t\rightarrow T$ in $f^{\gtrless}$ and ${\cal U}^{\gtrless}$). In this case, we obtain
\begin{eqnarray}
&&\left[\frac{\partial }{\partial T}+\frac{i}{\hbar}\left(\varepsilon({\bm k}+\frac{{\bm\kappa}}{2})
-\varepsilon({\bm k}-\frac{{\bm\kappa}}{2})\right)-\frac{\partial {\bm A}(T,t)}{\partial t}\nabla_{\bm k}
+\frac{\partial {\bm A}(T,t)}{\partial T}\nabla_{\bm\kappa}\right]
f^{<}({\bm k},{\bm\kappa}|T,t)\nonumber\\
&&=\sum\limits_{{\bm k}_1}\int\limits_{-\infty}^{\infty}dt_1
f^{<}({\bm k}_1,{\bm\kappa}|T,t-t_1)
W({\bm k}_1,{\bm k},{\bm\kappa}|T,t,t_1),
\label{kin1}
\end{eqnarray}
where the scattering probability $W$ is easily read off from Eqs.~(\ref{Dyson3}) to (\ref{FF2}). For a better readability, the result is presented in the Appendix. This kinetic equation still preserves the double-time nature of GFs, but restricts to locality in time with respect to the macroscopic time scale $T$. From Eq.~(\ref{kin1}), an important sum rule for ${\bm\kappa}={\bm 0}$ follows. Calculating the sum over ${\bm k}$, we immediately obtain from the left hand side of Eq.~(\ref{kin1}) the equality
\begin{equation}
\sum\limits_{\bm k}W({\bm k}_1,{\bm k},{\bm\kappa}={\bm 0}|T,t,t_1)=0,
\label{sumrule}
\end{equation}
which is confirmed from Eqs.~(\ref{Dyson3}) to (\ref{FF2}) by taking into account the symmetry relation
\begin{equation}
U^{<}_{\lambda}({\bm k},{\bm k}^{\prime},{\bm\kappa}|T,t)=
U^{>}_{\lambda}({\bm k}^{\prime},{\bm k},-{\bm\kappa}|T,-t).
\label{symU}
\end{equation}
The sum rule in Eq.~(\ref{sumrule}) (and its multi-band extension) plays an important role in the kinetic transport theory.

The kinetic description further simplifies, when the double-time character of the problem is completely neglected: $f^{\gtrless}(T,t)\rightarrow f^{\gtrless}(T)$. Specializing to electron-phonon interaction of the Fr\"ohlich type and restricting to ${\bm\kappa}={\bm 0}$, we obtain for the scattering probability
\begin{eqnarray}
W({\bm k}^{\prime},{\bm k}|T)&&=2{\rm Re}\sum\limits_{{\bm q},\lambda}
\int\limits_{0}^{\infty}dt_1e^{-st_1}D_{\lambda}^{>}({\bm q}|t_1)
f^{>}({\bm k}^{\prime}+{\bm q}|T)\label{scatW}\\
&&\times\left[
P({\bm k}^{\prime}+\frac{\bm q}{2},{\bm k}-\frac{\bm q}{2},{\bm q}|T,t_1)
-P({\bm k}^{\prime}+\frac{\bm q}{2},{\bm k}+\frac{\bm q}{2},{\bm q}|T,t_1)
\right],\nonumber
\end{eqnarray}
with the following field-dependent phase factor
\begin{eqnarray}
P({\bm k}^{\prime},{\bm k},{\bm q}|T,t_1)&&=
\exp\Biggl\{
\frac{i}{\hbar}\int\limits_{0}^{t_1}d\tau\left[
\varepsilon({\bm k}^{\prime}+\frac{\bm q}{2}+
\int\limits_{T-t_1}^{T-t_1+\tau}d\tau^{\prime}{\bm F}(\tau^{\prime}))\right.\nonumber\\
&&\left.
-\varepsilon({\bm k}^{\prime}-\frac{\bm q}{2}+
\int\limits_{T-t_1}^{T-t_1+\tau}d\tau^{\prime}{\bm F}(\tau^{\prime}))
\right]\Biggl\}
\delta_{{\bm k}^{\prime},{\bm k}+\int\limits_{T}^{T-t_1}d\tau {\bm F}(\tau)}.
\label{PP}
\end{eqnarray}
Here, $s$ denotes the Laplace variable of the rudimental microscopic time variation that regularizes the $t_1$ integral, and ${\bm F}(\tau)$ is an abbreviation for $e{\bm E}(\tau)/\hbar$. From Eq.~(\ref{scatW}), the sum rule in Eq.~(\ref{sumrule}) is again easily verified. Finally, we obtain the kinetic equation for the transport under the influence of a constant electric field (${\bm E}(t)\rightarrow {\bm E}$) in a form that was published many years ago \cite{Bryksin:2373} (cf., also Ref. [\onlinecite{Haug}]).

\subsection{Current density}
An expression for the current density is naturally derived from the general conservation law for the particle number. To illustrate the procedure, let us treat the kinetic equation for particles with kinetic energy $\varepsilon({\bm k})$ that are scattered via the Coulomb interaction. The conservation law is easily expressed by nonequilibrium GFs (c.f, for instance Ref. [\onlinecite{Schindl:235106}]), and a straightforward calculation leads to the following result expressed in the momentum representation
\begin{equation}
{\bm j}(t)=en\sum\limits_{\bm k}\frac{1}{\hbar}\nabla_{\bm k}\varepsilon({\bm k})f^{<}({\bm k},{\bm\kappa}={\bm 0}|t,0). \label{conventional}
\end{equation}
Scattering does not explicitly enter this equation, in which $n$ denotes the carrier density. It is sufficient to calculate the distribution function $f^{<}({\bm k}|T,t=0)=f^{<}({\bm k},{\bm\kappa}={\bm 0}|T,t=0)$, which does not depend on the momentum ${\bm\kappa}$. How general is this conclusion? Obviously, hopping transport is not captured by Eq.~(\ref{conventional}) as localized states have no dispersion. In fact, Eq.~(\ref{conventional}) is a special result applicable to the well-studied models that can be integrated into a more general definition of the current density, which covers more complex systems with higher-order scattering. According to this definition, the current density is expressed by the time derivative of the dipole operator ${\bm j}(t)=(1/V)d{\bm D}/dt$. This physically appealing approach is indeed more general than Eq.~(\ref{conventional}). In the momentum representation, we have
\begin{equation}
{\bm j}(T)=ie\sum\limits_{\bm k}\nabla_{\bm\kappa}\left.
\frac{\partial}{\partial T}f^{<}({\bm k},{\bm\kappa}|T,t=0)\right|_{{\bm\kappa}={\bm 0}},
\label{stromalg}
\end{equation}
which is converted into another equivalent form by taking into account the kinetic Eq.~(\ref{kin1})
\begin{equation}
{\bm j}(T)=en\sum\limits_{\bm k}{\bm v}_{eff}({\bm k})f^{<}({\bm k}|T,0),
\label{strom}
\end{equation}
with an effective velocity given by
\begin{equation}
{\bm v}_{eff}({\bm k})={\bm v}({\bm k})+i\int\limits_{-\infty}^{\infty}dt_1
\frac{f^{<}({\bm k}|T,-t_1)}{f^{<}({\bm k}|T,0)}
\sum\limits_{{\bm k}^{\prime}}{\bm W}_1({\bm k},{\bm k}^{\prime}|T,t_1).
\label{veff}
\end{equation}
The drift velocity is denoted by $v({\bm k})=\nabla_{\bm k}\varepsilon({\bm k})/\hbar$, and the vector ${\bm W}_1({\bm k},{\bm k}^{\prime}|T,t_1)$ is an abbreviation for $\nabla_{\bm\kappa}W({\bm k},{\bm k}^{\prime},{\bm\kappa}|T,t=0,t_1)|_{{\bm\kappa}={\bm 0}}$. Whenever the sum rule for this vector field $\sum_{\bm k}{\bm W}_1({\bm k},{\bm k}^{\prime}|T,t_1)={\bm 0}$ is satisfied, Eqs.~(\ref{strom}) and (\ref{veff}) reproduce the conventional result given in Eq.~(\ref{conventional}). This fortunate situation happens, for instance, for the Fr\"ohlich electron-phonon coupling and the Coulomb interaction. In general, Eq.~(\ref{conventional}) is applicable, when the interaction Hamiltonian commutes with the dipole operator. However, this condition is not always fulfilled. For instance, for the transport of small polarons, the ${\bm W}_1$ contribution in Eq.~(\ref{veff}) is most essential so that only Eq.~(\ref{stromalg}) [or the equivalent Eqs.~(\ref{strom}) and (\ref{veff})] provides meaningful results. The definition of the current density in Eq.~(\ref{stromalg}) includes the ${\bm\kappa}$ gradient of the full distribution function $f^{<}({\bm k},{\bm\kappa}|T,t=0)$ at ${\bm\kappa}={\bm 0}$. This ${\bm\kappa}$ dependence reappears in Eq.~(\ref{veff}) via the vector ${\bm W}_1$. Consequently, it is not sufficient to deal with a distribution function that depends only on one quasi-momentum ${\bm k}$. In fact, the general basis for treating carrier transport is provided by the kinetic Eq.~(\ref{Dyson3}), from which the ${\bm\kappa}$ dependence can be determined.

In summary, we conclude that Eqs.~(\ref{strom}) and (\ref{veff}) put our former semi-phenomenological approach \cite{Bryksin:1415} on a firm microscopic basis and lead to a general expression for the current density that takes into account the two-time character of quantum transport.

To illustrate the additional ability of the approach to simultaneously cover transport via localized and extended states, let us, for simplicity, treat the steady-state transport in a one-time approximation under the influence of an applied electric field $E_{dc}$. The momentum representation in Eqs.~(\ref{strom}) and (\ref{veff}) is adapted to the description of low-field transport, when the states remain essentially extended. With increasing field strength due to Wannier-Stark (WS) localization, negative differential conductivity can appear. To describe this transport regime in a more appropriate fashion, the expression for the current density in Eqs.~(\ref{strom}) and (\ref{veff}) is rewritten in an exact manner. \cite{Bryksin:1415} Up to intracollisional field effects, the result for the current density along the direction of the electric field
\begin{equation}
j=-\frac{n}{E_{dc}}\sum_{{\bm k},{\bm k}^{\prime}}
(\varepsilon({\bm k})-\varepsilon({\bm k}^{\prime}))
f^{<}({\bm k}^{\prime})W({\bm k}^{\prime},{\bm k})
\label{stromhalb}
\end{equation}
is compatible with negative differential conductivity $j\sim 1/E_{dc}$. Moreover, Eq.~(\ref{stromhalb}) proves that there is no current in the absence of any inelastic scattering. A strong electric field generates Bloch oscillations that localize carriers so that there is no current as long as only elastic scattering is present. By switching to the Houston representation, one arrives at another equivalent expression \cite{Bryksin:1415}
\begin{equation}
j=en\sum\limits_{{\bm k}_{\perp},{\bm k}^{\prime}_{\perp}}\sum\limits_{l=-\infty}^{\infty}(ld)f^{<}({\bm k}_{\perp}^{\prime})W_{0l}^{0l}({\bm k}_{\perp}^{\prime},{\bm k}_{\perp}),
\label{stromhop}
\end{equation}
which clearly reveals the hopping character of the transport. $d$ denotes the periodicity of the lattice ($ld$ is the hopping length), and ${\bm k}_{\perp}$ is the momentum perpendicular to the field. The $l$-sum extends over the whole WS ladder. The potential of the approach is illustrated by its ability to unify the band and hopping picture by an exact reformulation that mediates between them.

\subsection{Diffusion coefficient}
What favors our approach to quantum diffusion is its close relationship to the carrier drift treated in the previous Subsection. Let us follow the same line of reasoning by first focusing on the regular part of the diffusion coefficient \cite{Bryksin:1415} defined by
\begin{equation}
D_0(t)=\frac{1}{2}\int d^3{\bm r}{\bm r}^2\frac{\partial}{\partial t}\langle\psi^{\dag}(x)\psi(x)\rangle,
\end{equation}
with $x=({\bm r},t)$ (the spin variable is not indicated). The equivalent expression in the momentum representation has a form that is similar to Eq.~(\ref{stromalg})
\begin{equation}
D_0(T)=\frac{i}{2}\sum\limits_{\bm k}\nabla_{\bm\kappa}^2\left.
\frac{\partial}{\partial T}G^{<}({\bm k},{\bm\kappa}|T,t=0)\right|_{{\bm\kappa}={\bm 0}}.
\label{diffdef}
\end{equation}
Again, we mention that both wave vectors ${\bm k}$ and ${\bm\kappa}$ appear in this definition. Therefore, the basic kinetic equation that describes quantum diffusion has to be formulated for the GF $f^{<}({\bm k},{\bm\kappa}|T,t)$, which comprises not only ${\bm k}$ but also ${\bm\kappa}$. Introducing the vector field
\begin{equation}
{\bm g}({\bm k}|T,t)=i{\bm f}_1^{<}({\bm k}|T,t)\equiv i\nabla_{\bm\kappa}f^{<}({\bm k},{\bm\kappa}|T,t)|_{{\bm\kappa}={\bm 0}},
\label{defg}
\end{equation}
we obtain the equivalent form
\begin{eqnarray}
D_0(T)&&=\sum\limits_{\bm k}\left\{
{\bm v}({\bm k})\cdot{\bm g}({\bm k}|T,0)
+i\int\limits_{-\infty}^{\infty}dt_1{\bm g}({\bm k}|T,-t_1)\cdot
\sum\limits_{{\bm k}^{\prime}}{\bm W}_1({\bm k},{\bm k}^{\prime}|T,t_1)
\right\}\nonumber\\
&&-\frac{1}{2}\sum\limits_{\bm k}\int\limits_{-\infty}^{\infty}dt_1
f^{<}({\bm k}|T,-t_1)
\sum\limits_{{\bm k}^{\prime}}{W}_2({\bm k},{\bm k}^{\prime}|T,t_1),
\label{eqD0}
\end{eqnarray}
in which $W_2$ denotes the second derivative $\nabla_{\bm\kappa}^{2} W|_{{\bm\kappa}={\bm 0}}$. The explicit scattering contributions in Eq.~(\ref{eqD0}) indicated by ${\bm W}_1$ and $W_2$ vanish for widespread models such as the Fr\"ohlich electron-phonon coupling and the Coulomb interaction. In contrast to the current density, which is governed by the distribution function $f^{<}$, diffusion phenomena are described by means of the vector ${\bm g}$, which satisfies its own kinetic equation that is easily obtained from Eqs.~(\ref{kin1}) and (\ref{defg}). This procedure unambiguously determines the contribution $D_0(T)$, in which, however, the irregular part is still missing. To complete the calculation of the total diffusion coefficient, the vector ${\bm g}$ in Eq.~(\ref{eqD0}) is replaced by a new quantity ${\bm\varphi}$ that solves the same kinetic equation as ${\bm g}$, but with a modified inhomogeneity that is compatible with the constraint
\begin{equation}
\sum\limits_{{\bm k}}{\bm\varphi}({\bm k}|T,t)={\bm 0}.
\label{constraint}
\end{equation}
Accordingly, the basic quantity ${\bm\varphi}$, which determines quantum diffusion via the diffusion coefficient $D(T)$, satisfies the quantum-kinetic equation
\begin{eqnarray}
&&\left[\frac{\partial}{\partial T}-\frac{\partial{\bm A}(T,t)}{\partial t}\nabla_{\bm k} \right]
{\bm\varphi}({\bm k}|T,t)=
\sum\limits_{{\bm k}_1}\int\limits_{-\infty}^{\infty}dt_1{\bm\varphi}({\bm k}_1|T,t-t_1)
W({\bm k}_1,{\bm k}|T,t,t_1)\nonumber\\
&&+{\bm v}({\bm k})f^{<}({\bm k}|T,t)-\sum\limits_{{\bm k}_1}
{\bm v}({\bm k}_1)f^{<}({\bm k}_1|T,t)\label{eqphi}\\
&&+i\sum\limits_{{\bm k}_1}\int\limits_{-\infty}^{\infty}dt_1
f^{<}({\bm k}_1|T,t-t_1)\left[{\bm W}_1({\bm k}_1,{\bm k}|T,t,t_1)
-\sum\limits_{{\bm k}_2}{\bm W}_1({\bm k}_1,{\bm k}_2|T,t,t_1) \right],\nonumber
\end{eqnarray}
which is clearly in line with the sum rule in Eq.~(\ref{constraint}).

To familiarize oneself with the derivation of basic results concerning quantum diffusion, let us work with the Carson-Heaviside transformation of the kinetic equation with respect to the time variable $T$ [$f(s)=s\int_{0}^{\infty}dT\exp(-sT)f(T)$] in the treatment of a constant electric field. From the kinetic equation for ${\bm g}$ and the sum rule in Eq.~(\ref{constraint}), we obtain
\begin{equation}
{\bm g}({\bm k}|s,t)={\bm\varphi}({\bm k}| s,t)+\frac{1}{s}{\bm B}(s,t)f^{<}({\bm k}|s,t),
\label{gandphi}
\end{equation}
in which the following quantities appear
\begin{equation}
{\bm B}(s,t)=\sum\limits_{\bm k}{\bm v}_{eff}({\bm k}|s,t)f^{<}({\bm k}|s,t),
\end{equation}
\begin{equation}
{\bm v}_{eff}({\bm k}|s,t)={\bm v}({\bm k})+i\int\limits_{-\infty}^{\infty}dt_1\frac{f^{<}({\bm k}|s,t-t_1)}{f^{<}({\bm k}|s,t)}\sum\limits_{{\bm k}^{\prime}}{\bm W}_1({\bm k},{\bm k}^{\prime}|t,t_1).
\end{equation}
Replacing the vector ${\bm g}$ in Eq.~(\ref{eqD0}) by the new vector field ${\bm\varphi}$ according to Eq.~(\ref{gandphi}), we obtain
\begin{eqnarray}
D_0(s)&&=\sum\limits_{\bm k}\left\{
{\bm v}({\bm k})\cdot{\bm\varphi}({\bm k}|s,0)
+i\int\limits_{-\infty}^{\infty}dt_1{\bm\varphi}({\bm k}|s,-t_1)\cdot
\sum\limits_{{\bm k}^{\prime}}{\bm W}_1({\bm k},{\bm k}^{\prime}|0,t_1)
\right\}\nonumber\\
&&-\frac{1}{2}\sum\limits_{\bm k}\int\limits_{-\infty}^{\infty}dt_1
f^{<}({\bm k}|s,-t_1)
\sum\limits_{{\bm k}^{\prime}}{W}_2({\bm k},{\bm k}^{\prime}|0,t_1)\nonumber\\
&&+\frac{1}{s}\biggl\{\sum\limits_{\bm k}{\bm v}_{eff}({\bm k}|s,0)f^{<}({\bm k}|s,0) \biggl\}^2.
\label{eqD1}
\end{eqnarray}
The last term on the RHS of this equation is nothing but the irregular contribution, which is subtracted out according to the proper definition of the diffusion coefficient. \cite{Bryksin:1415} Consequently, only the first three terms on the RHS of Eq.~(\ref{eqD1}) survive and define the total diffusion coefficient $D(s)$. These results provide a rigorous theory of quantum diffusion, in which the double-time character is accounted for by a vector field ${\bm\varphi}$ that is the solution of the specific quantum-kinetic Eq.~(\ref{eqphi}). At this stage, the theories of quantum transport and quantum diffusion have reached the same level of sophistication.

The above theory of quantum diffusion is formulated in the momentum representation, which is adapted to extended states. By exact manipulations, other equations for the diffusion coefficient are obtained that are more appropriate in the WS regime, when carriers execute Bloch oscillations. Within perturbation theory with respect to scattering, for which the Hamiltonian commutes with the dipole operator, we obtain for the steady state
\begin{equation}
D_{zz}=\sum\limits_{\bm k}v_z({\bm k})\varphi({\bm k})
=\frac{1}{2(eE_{dc})^2}\sum\limits_{{\bm k},{\bm k}^{\prime}}\left(\varepsilon(k_z)-\varepsilon(k_z^{\prime}) \right)^2 f^{<}({\bm k}^{\prime})W({\bm k}^{\prime},{\bm k}),
\label{locD}
\end{equation}
which has the same structure as Eq.~(\ref{stromhalb}) derived in the previous Subsection. In the targeted regime of field-induced localization, there is no diffusion without inelastic scattering.

A more general expression for the diffusion coefficient applicable to the WS regime is derived within the outlined approach by exploiting the WS representation. \cite{Bryksin:1415} The final result
\begin{equation}
D=\frac{1}{2}\sum\limits_{{\bm k}_{\perp},{\bm k}_{\perp}^{\prime}}\sum\limits_{l=-\infty}^{\infty}
(ld)^2f^{<}({\bm k}_{\perp}^{\prime})\widetilde{W}_{0,l}^{0,l}({\bm k}_{\perp}^{\prime},{\bm k}_{\perp}),
\end{equation}
with $\widetilde{W}$ being an effective scattering probability, allows an interpretation within the hopping picture that relates carrier diffusion to the square of the hopping length $(ld)^2$, the lateral carrier distribution function $f^{<}({\bm k}_{\perp})$, and the scattering probability in the site representation.

\section{Example: Phononless transport}
Based on the nonequilibrium GF technique, a unified approach has been developed that covers both quantum transport and quantum diffusion and that is likewise applicable to transport via extended and localized states. A salient feature of this theory is the double-time character of quantum transport. A natural question arises: What is the significance of this double-time dependence? An answer is gained only beyond the generalized Kadanoff-Baym ansatz. Generally speaking, it is difficult to draw an overall conclusion regarding the physical potential of the two-time quantum kinetics. Summarizing the bulk of conventional transport studies, it is tempting to assume that the double-time approach resolves only minor corrections that are more or less unimportant. That this assessment cannot be the whole truth will be illustrated by a macroscopic transport phenomenon that has no analogy in the conventional approach because of its strict double-time character. To be more specific, a steady-state current will be identified in the WS regime that is driven by dc and ac electric fields without the participation of any inelastic scattering. This phononless current appears only beyond the Kadanoff-Baym ansatz and is due to the double-time dependence of the GFs.

\subsection{Solution of the kinetic equation}
The double-time dependence is studied by a model calculation that is simple enough to allow for an analytical solution. Some results obtained by a ${\bm\kappa}$-independent approach have already been published previously. \cite{Bryksin:1235} The model refers to a one-dimensional semiconductor superlattice, which is biased by dc and ac electric fields
\begin{equation}
E(t)=E_{dc}+E_{ac}\cos(\omega_{ac}t),
\end{equation}
that are sufficiently strong so that WS localization occurs ($\Omega_{dc}\tau \gg 1$ and $\Omega_{ac}\tau\gg 1$, with $\tau$ being an effective scattering time and $\Omega_{ac,dc}=eE_{ac,dc}d/\hbar$). Bloch oscillations that appear in this transport regime are accounted for by a discrete Fourier transformation of the GFs
\begin{equation}
f^{\gtrless}(k|T,t)=\sum\limits_{l=-\infty}^{\infty}
f_{l}^{\gtrless}(T,t)e^{ilkd}.
\end{equation}
To calculate the Fourier components of the double-time distribution function $f_{l}^{<}(T,t)$, we treat scattering on polar optical phonons with energy $\hbar\omega_0$ and neglect the smooth $q$ dependence of the coupling [$D_{\lambda}^{\gtrless}(q|t)\rightarrow D^{\gtrless}(t)$]. Within the WS regime, only the $l=0$ component of the GFs $f_{l}^{\gtrless}$ enter the collision integral so that Eq.~(\ref{Dyson3}) takes the form
\begin{eqnarray}
&&\left\{\frac{\partial}{\partial T}+il\Omega_{dc}+il\Omega_{ac}\cos(\omega_{ac}T)
\cos\left(\frac{\omega_{ac}t}{2} \right) \right\}f_{l}^{<}(T,t)=
\sum\limits_{k,q}e^{-ilkd}\int\limits_{0}^{\infty}dt_1\label{kin2}\\
&&\times\biggl\{
\left[D^{<}(t-t_1)\Phi_{t,t_1}(k,q)-D^{>}(t_1)\Phi_{t,t_1}^{*}(k,q) \right]
f_{0}^{<}(T-\frac{t_1}{2},t-t_1)f_{0}^{>}(T+\frac{t-t_1}{2},t_1)\nonumber\\
&&-\left[D^{>}(-t_1)\Phi_{t_1,0}^{*}(k,q)-D^{<}(t+t_1)\Phi_{t_1,0}(k,q) \right]
f_{0}^{<}(T-\frac{t_1}{2},t+t_1)f_{0}^{>}(T-\frac{t+t_1}{2},-t_1)
\biggl\},\nonumber
\end{eqnarray}
with the following field-dependent phase factor
\begin{eqnarray}
\Phi_{t,t_1}(k,q)&&=\exp\biggl\{
\frac{i}{\hbar}\int\limits_{t_1}^{t}d\tau\left[
\varepsilon\left(k+ q+A(\tau+T-\frac{t}{2}-t_1)-A(T-\frac{t}{2}) \right)
\right.\\
&&\left. -\varepsilon\left(k+A(\tau+T-\frac{t}{2}-t_1)-A(T-\frac{t}{2}) \right)
\right]\biggl\}.
\end{eqnarray}
The double-time character of the approach is still present in Eq.~(\ref{kin2}). The non-Markovian behavior extends both over the macroscopic ($T$) and microscopic ($t$) time scale. The main source of the $T$ dependence is the ac electric field that appears directly on the left hand side of Eq.~(\ref{kin2}). In most approaches, the double-time dependence is neglected by omitting the $t$ dependence in $f_{l}^{\gtrless}(T,t)$, which results from microscopic scattering processes described by the RHS of Eq.~(\ref{kin2}). However, both different lines of time evolution are generally coupled to each other by a convolution integral, the field-dependent kernel of which determines the role played by the kinetic history. An analytic solution of Eq.~(\ref{kin2}) is found for weakly coupled superlattices with the dispersion relation
\begin{equation}
\varepsilon(k)=\frac{\Delta}{2}(1-\cos(kd)).
\end{equation}
Considering the periodicity with respect to the $T$ dependence
\begin{equation}
f_{l}^{\gtrless}(T+2\pi/\omega_{ac},t)=f_{l}^{\gtrless}(T,t),\quad
f_{l}^{\gtrless}(T,t)=\sum\limits_{m=-\infty}^{\infty}
f_{l}^{\gtrless}(m,t)e^{im\omega_{ac}T},
\end{equation}
the main Fourier component $f_0^{<}(m=0,t)$ of the WS regime ($\Omega_{ac}\gg 1$) is calculated from the homogeneous integral equation
\begin{equation}
\int\limits_{-\infty}^{\infty}dt_1\left[
D^{<}(t_1)f_{0}^{<}(0,t_1)f_{0}^{>}(0,t-t_1)
-D^{>}(t_1)f_{0}^{>}(0,t_1)f_{0}^{<}(0,t-t_1)
\right]=0.
\label{kin3}
\end{equation}
This equation, derived under the condition of narrow minibands $\Delta/\hbar\Omega_{ac,dc}\ll 1$, determines the time dependence of the distribution function that appears beyond the generalized Kadanoff-Baym ansatz and that was ignored in most previous approaches. A solution is searched for in Fourier space by adopting the ansatz
\begin{equation}
f_{0}^{<}(0,\omega)=f_{0}^{>}(0,\omega)f(\omega),
\label{deff}
\end{equation}
and by considering the normalization condition
\begin{equation}
\int\limits_{-\infty}^{\infty}\frac{d\omega}{2\pi}f_{0}^{>}(0,\omega)f(\omega)=1.
\label{norm}
\end{equation}
Inserting the expressions
\begin{equation}
D^{\gtrless}(\omega)=\frac{\gamma_0\pi}{\sinh(\beta/2)}
\left[\delta(\omega+\omega_0)e^{\pm \beta/2}+
\delta(\omega-\omega_0)e^{\mp \beta/2} \right]
\label{elphon}
\end{equation}
for the electron-phonon coupling, it is easily verified that an exponential function in $\omega$ solves the Fourier-transformed version of Eq.~(\ref{kin3}). To determine the prefactor from Eq.~(\ref{norm}), the function $f_0^{>}(0,\omega)$ is needed. In our previous studies, \cite{Bryksin:233,Bryksin:1235} we obtained for weakly coupled superlattices ($\Delta\rightarrow 0$)
\begin{equation}
f_{0}^{>}(0,\omega)=\left\{ \begin{array}{c}
                              \frac{1}{2\pi U}\sqrt{4U-\omega^2},\quad |\omega|<2\sqrt{U} \\
                              0,\quad {\rm otherwise}
                            \end{array} \right. ,
\label{Ggros}
\end{equation}
with $U$ denoting the coupling strength of white-noise elastic scattering on impurities. The most remarkable features of this density of states function are its nonanalytic character with respect to the coupling $U$ and the absence of tails at the band edges. From Eqs.~(\ref{norm}) and (\ref{Ggros}), we obtain the final result for the distribution function
\begin{equation}
f(\omega)=\frac{\beta_{u}}{I_1(\beta_{u})}\exp\left(\frac{\hbar\omega}{k_BT} \right),\quad
\beta_{u}=\frac{2\hbar\sqrt{U}}{k_BT},
\label{distfunc}
\end{equation}
which applies whenever carriers thermalize more quickly in a given quantum well than they need to escape by tunneling. $I_1$ denotes the modified Bessel function. With increasing miniband width $\Delta$, the layers are more strongly coupled to each other, and the solution in Eq.~(\ref{distfunc}) is no longer adequate. In this case, only a numerical solution of Eq.~(\ref{kin3}) is available that accounts for a non-Markovian time evolution.

\subsection{Current density and diffusion coefficient}
Despite previous results concerning the hopping transport in the WS regime, let us look for a phononless transport mechanism by exploiting the more general two-time approach. The amazing result will be that there is in fact a phononless transport, when the double-time dependence of the GFs is properly accounted for.

Let us first focus on the constant steady-state current that is driven by external dc and ac electric fields under the exclusive influence of short-range elastic scattering on impurities. The steady-state current density
\begin{eqnarray}
j&&=en\sum\limits_{k}\frac{1}{\hbar}\frac{d\varepsilon(k)}{d k}\frac{\omega_{ac}}{2\pi}
\int\limits_{0}^{2\pi/\omega_{ac}}dTf^{<}(k|T,t=0)\label{strommodel}\\
&&=en\frac{\Delta d}{2\hbar}\frac{1}{2i}\left[f_{l=-1}^{<}(m=0,t=0)-f_{l=1}^{<}(m=0,t=0) \right],
\nonumber
\end{eqnarray}
is expressed by the components $f_{l=\pm 1}^{<}(0,0)$ of the distribution function that according to Eq.~(\ref{Dyson3}) obey the kinetic equation
\begin{eqnarray}
&&\biggl\{\frac{\partial}{\partial T}+il\Omega_{dc}+il\Omega_{ac}\cos(\omega_{ac}T) \biggl\}
f_{l}^{<}(T,0)\label{kin4}\\
&&=U\sum\limits_{k,q}e^{-ilkd}\int\limits_{0}^{\infty}dt_1\left[
\Phi_{0t_1}(k,q)-\Phi_{0t_1}^{*}(k,q) \right]\nonumber\\
&&\times\biggl\{f_0^{<}(T-\frac{t_1}{2},-t_1)f_0^{>}(T-\frac{t_1}{2},t_1)
-f_0^{<}(T-\frac{t_1}{2},t_1)f_0^{>}(T-\frac{t_1}{2},-t_1) \biggl\}
\equiv P_{l}(T).\nonumber
\end{eqnarray}
It is a consequence of this equation that no current can flow through the superlattice, when the variations on the microscopic time scale disappear: $f_0^{\gtrless}(T,t)\rightarrow f_0^{\gtrless}(T)$. This fact confirms the general conclusion mentioned in previous Sections that within the one-time picture only inelastic scattering enables carrier transport in the WS regime. However, the double-time nature of the kinetic evolution opens up a new channel, which enables phononless transport of Bloch oscillating carriers. This specific transport mechanism appears only beyond the generalized Kadanoff-Baym ansatz.

The formal solution of Eq.~(\ref{kin4}) has the form
\begin{equation}
f_{l}^{<}(0,0)=\sum\limits_{m=-\infty}^{\infty}\frac{P_{l,m}S_{l,m}}{il\Omega_{dc}},\quad
S_{l,m}=\sum\limits_{k=-\infty}^{\infty}J_{k-m}\left(l\frac{\Omega_{ac}}{\omega_{ac}} \right)
J_{k}\left(l\frac{\Omega_{ac}}{\omega_{ac}} \right)
\frac{l\Omega_{dc}}{l\Omega_{dc}+k\omega_{ac}},
\label{formalsol}
\end{equation}
where $P_{l,m}$ are the Fourier components of $P_{l}(T)$, the calculation of which follows the same steps as outlined in the previous Section. A straightforward procedure applicable to weakly coupled superlattices ($\Delta\rightarrow 0$) leads to the final result for the current density in steady state
\begin{eqnarray}
&&j=env_d,\quad
v_d=\frac{\pi U\Delta^2d}{8\hbar^2}\frac{J_{\nu}(\nu^{\prime})J_{-\nu}(\nu^{\prime})}{\omega_{ac}^2\sin(\pi\nu)}
\sum\limits_{k}\frac{J_k^2(\nu^{\prime})}{k+\nu}\label{finalcurrent}\\
&&\times\int\limits_{-\infty}^{\infty}\frac{d\omega}{2\pi}f_{0}^{>}(0,\omega)
f_{0}^{>}(0,\omega+\omega_{ac}(k+\nu))
\left[f(\omega+\omega_{ac}(k+\nu))-f(\omega) \right],
\nonumber
\end{eqnarray}
with the abbreviations $\nu=\Omega_{dc}/\omega_{ac}$ and $\nu^{\prime}=\Omega_{ac}/\omega_{ac}$. This result again confirms that there is no phononless current, when the trivial solution $f(\omega)=1$ is accepted, which is suggested by the Kadanoff-Baym ansatz within a strict one-time approach. The specific constant current contribution in Eq.~(\ref{finalcurrent}) disappears also, when the ac field is switched off ($\Omega_{ac}=0$) because the combined density of states vanishes for $2\Omega_{dc}>\sqrt{U}$.
\begin{figure}
\centerline{\epsfig{file=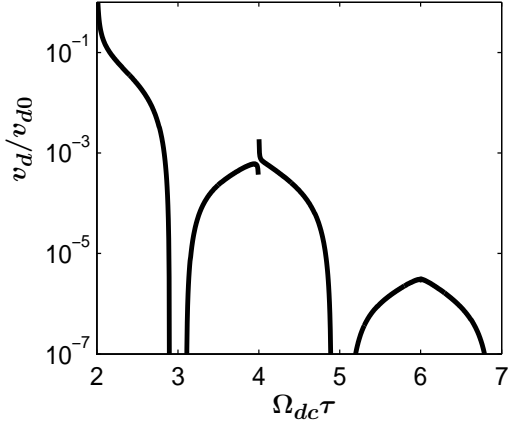,width=7.0cm}}
\caption{The normalized drift velocity $v_d$ calculated from Eq.~(\ref{finalcurrent}) as a function of $\Omega_{dc}\tau$ for $\beta_U=3$, $\omega_{ac}\tau=2$, $\Omega_{ac}\tau=1$, and $U\tau^2=0.05$. The quantity $v_{d0}$ is given by $\Delta^2\tau d/(35\pi \hbar^2)$, and $\tau$ denotes the scattering time.}
\label{Figure1}
\end{figure}
Consequently, the assertion that there is no current in the hopping regime without any inelastic scattering remains valid also in the two-time approach, when only a constant electric field is applied.

The calculated drift velocity is shown in Fig.~1 as a function of the dc electric field. Photon replicas centered around $\Omega_{dc}=k\omega_{ac}$ are separated by gaps that result from the edge structure of the combined density of states. As a drawback of the simplified treatment of scattering, singularities appear in the current-voltage characteristics at $\Omega_{dc}=k\omega_{ac}$. A damping of these resonances is easily accounted for in more refined and realistic approaches, which unlikely alter the qualitative physics discussed in this Section.

The diffusion coefficient is obtained by a similar calculation. According to Eq.~(\ref{eqD1}), $D$ is expressed by the function $\varphi$
\begin{equation}
D=\sum\limits_{k}\frac{1}{\hbar}\frac{d\varepsilon(k)}{dk}\frac{\omega_{ac}}{2\pi}
\int\limits_{0}^{2\pi/\omega_{ac}}dT\varphi(k|T,t=0),
\end{equation}
which satisfies the following quantum-kinetic equation
\begin{eqnarray}
&&\biggl\{\frac{\partial}{\partial T}+il\Omega_{dc}+il\Omega_{ac}\cos(\omega_{ac}T)\cos\left(\frac{\omega_{ac}t}{2} \right) \biggl\}
\varphi_{l}(T,t)\label{kinphil}\\
&&=i\frac{\Delta d}{4\hbar}\left\{f_{l+1}^{<}(T,t)-f_{l-1}^{<}(T,t)-(f_1^{<}(T,t)-f_{-1}^{<}(T,t)) \right\}\nonumber\\
&&+\frac{d}{2\pi}\int\limits_0^{2\pi/d}dke^{-ilkd}\sum\limits_{k_1}\int\limits_{-\infty}^{\infty}dt_1
\varphi(k_1|T,t-t_1)W(k_1,k|T,t,t_1)
\equiv P_{l}(T).\nonumber
\end{eqnarray}
This equation differs from Eq.~(\ref{kin4}) for $f_{l}^{<}(T,t)$ by the appearance of a inhomogeneous term. The formal solution has the form of Eq.~(\ref{formalsol}). Restricting the calculation of $P_{l,m}(t=0)$ to its lowest-order contribution, we obtain the final result for the diffusion coefficient applicable to the WS regime
\begin{equation}
D=\frac{\Delta}{\hbar}\sum\limits_{k=-\infty}^{\infty}\frac{J_k^2\left({\Omega_{ac}}/{\omega_{ac}} \right)}{\Omega_{dc}+k\omega_{ac}}\,
\frac{v_dd}{2},
\end{equation}
which can be formally interpreted in terms of a relationship suggested for the current density more than 40 years ago \cite{Tien:647}
\begin{equation}
D(\Omega_{dc},\Omega_{ac})=\sum\limits_{k=-\infty}^{\infty}
J_k^2\left(\frac{\Omega_{ac}}{\omega_{ac}} \right)
D(\Omega_{dc}+k\omega_{ac}).
\label{Tucker}
\end{equation}
Accordingly, the diffusion coefficient under the combined influence of ac and dc electric fields is easily obtained from the quantity
\begin{equation}
D(\Omega_{dc})=\frac{v_dd}{2}\coth\left(\frac{\hbar\Omega_{dc}}{\Delta} \right),
\end{equation}
which refers to the absence of the ac electric field $\Omega_{ac}=0$. However, this interpretation has only a formal character and heavily depends on the approximations made in the derivation. The main conclusion is the same as for the current density, namely that this kind of high-field quantum diffusion appears only beyond the Kadanoff-Baym ansatz by a strict treatment of the double-time dependence. The experimental demonstration of this exclusive double-time quantum effect should be feasible by studying biased quantum-box superlattices.

\section{Summary}
Starting from a semi-phenomenological kinetic approach, a unified one-electron theory of quantum transport and quantum diffusion was developed in previous works \cite{Bryksin:1415,Kleinert:315} that applies both to the band picture applicable to extended states at low electric fields and to the hopping picture for transport under quantizing electric field. Both approaches are completely equivalent and can be mutually derived from each other. For the current density, this equivalence was already demonstrated in Ref. [\onlinecite{Bryksin:2373}]. Furthermore, a comparative treatment of carrier drift and diffusion \cite{Bryksin:1415,Kleinert:315} revealed the particular nature of quantum diffusion. Whereas the drift velocity goes back to the nonequilibrium distribution function, the diffusion coefficient turns out to be constructed from a derived quantity that does not solve the Boltzmann equation or its quantum-kinetic extension. Most disturbing was the necessity to deal with the total GF $f^{<}({\bm k},{\bm\kappa}|T,t)$ that depends on two wave vectors ${\bm k}$ and ${\bm\kappa}$. The ${\bm\kappa}$ dependence seems to be dispensable for the description of transport in homogeneous systems, which are translational invariant. However, to probe carrier diffusion, at least an initial inhomogeneity of the carrier ensemble is necessary so that the ${\bm\kappa}$ dependence must be preserved in the unified description of drift and diffusion.

In this paper, we put the former semi-phenomenological approach on a firm microscopic basis by applying nonequilibrium GF techniques. The unified theory of quantum transport and quantum diffusion has been constructed from the quantum-kinetic equation for the full distribution function $f^{<}({\bm k},{\bm\kappa}|T,t)$. The most salient feature of this extension are the appearance of the double-time nature of quantum kinetics and the related non-Markovian evolution in two time channels, namely the microscopic and macroscopic time regime. On the fundamental microscopic level, quantum evolution seems to be more general than the classical schema that dictates a strictly one-dimensional progression from the past to the future. An interesting question concerns the relevance of the two-time quantum kinetics, namely whether it is possible that new physics appears in this domain. A preliminary answer provides the treatment of a one-dimensional superlattice subject to dc and ac electric fields. The existence of phononless carrier transport and diffusion is demonstrated, the origin of which is the two-time dependence of the GFs. This distinct steady-state transport mechanism appears only beyond the generalized Kadanoff-Baym ansatz. Its experimental verification seems to be feasible by studying quantum-box superlattices.

The rigorous two-time quantum-kinetic approach presented in this paper is likewise applicable to quantum transport and quantum diffusion and covers both transport via extended states and hopping between localized carriers.

\section*{Acknowledgements}
I very much acknowledge early stimulating discussions with Prof. Dr. Valerij V. Bryksin, who recently passed away.

\appendix
\section{Scattering probability in eq.~(\ref{kin1})}
Assuming strict locality in time $T$, we obtain from the kinetic Eq.~(\ref{Dyson3}) and the definitions in Eqs.~(\ref{eqP}) to (\ref{FF2})
\begin{eqnarray}
&&W({\bm k}_1,{\bm k},{\bm\kappa}|T,t,t_1)\label{defW}\\
&=&\sum\limits_{{\bm q},\lambda}\biggl\{
\Theta(t_1)f^{>}({\bm k}_1-{\bm q}-\frac{{\bm\kappa}}{2}|T,t_1)P_{t_1}({\bm k}_1,{\bm q},{\bm\kappa}|T,t)
\nonumber\\
&\times&\left[
\delta_{{\bm k}_1,{\bm k}+{\bm q}+{\bm A}(T+t/2-t_1)-{\bm A}(T-t/2)}
{\cal U}_{\lambda}^{<}({\bm k}_1-{\bm q},{\bm k}_1,{\bm\kappa}|T,t-t_1)\right.
\nonumber\\
&-&\left.
\delta_{{\bm k}_1,{\bm k}+{\bm A}(T+t/2-t_1)-{\bm A}(T-t/2)}
{\cal U}_{\lambda}^{>}({\bm k}_1-\frac{\bm\kappa}{2},{\bm k}_1-{\bm q}-\frac{\bm\kappa}{2},{\bm 0}|T,t_1)
\right]\nonumber\\
&-& \Theta(-t_1)f^{>}({\bm k}_1-{\bm q}+\frac{{\bm\kappa}}{2}|T,t_1)Q_{t_1}({\bm k}_1,{\bm q},{\bm\kappa}|T,t)
\nonumber\\
&\times&\left[
\delta_{{\bm k}_1,{\bm k}+{\bm A}(T-t/2+t_1)-{\bm A}(T-t/2)}
{\cal U}_{\lambda}^{>}({\bm k}_1+\frac{\bm\kappa}{2},{\bm k}_1-{\bm q}+\frac{\bm\kappa}{2},{\bm 0}|T,t_1)\right.
\nonumber\\
&-&\left.
\delta_{{\bm k}_1,{\bm k}+{\bm q}+{\bm A}(T-t/2+t_1)-{\bm A}(T-t/2)}
{\cal U}_{\lambda}^{<}({\bm k}_1-{\bm q},{\bm k}_1,{\bm\kappa}|T,t-t_1)
\right]
\biggl\}.\nonumber
\end{eqnarray}
This equation completes the two-time quantum kinetic Eq.~(\ref{kin1}).




\end{document}